\documentclass[10pt,leqno]{amsart}
\usepackage{graphicx}
\usepackage{indentfirst,csquotes}
\usepackage{xcolor}
\usepackage{amssymb,amsthm,amsmath}
\usepackage{xcolor,paralist,hyperref,titlesec,fancyhdr,etoolbox}

\titleformat{\section}[display]{\normalfont\huge\bfseries\centering}{\centering\chaptertitlename\thechapter}{10pt}{\Large}
\titlespacing*{\section}{0pt}{0ex}{0ex}

\hypersetup{ colorlinks=true, linkcolor=black, filecolor=black, urlcolor=black }

\usepackage{lipsum}

\begin{document}

\title{Current status of the Extension of the FRIPON network in Chile} 

\author[Initial Surname]{Felipe Gutiérrez Rojas$^1$, Sébastien Bouquillon$^{2,3}$, Rene A. Mendez$^4$, Hernan Pulgar$^4$, Marcelo Tala Pinto$^{5,6}$, Katherine Vieira$^7$, Millarca Valenzuela Picón$^8$, Andrés Jordán$^{5,6,9}$, Christian H.R. Nitschelm$^{10}$, Massinissa Hadjara$^{11}$, José Luis Nilo Castellón$^{12}$, Maja Vuckovic$^{13}$, Hebe Cremades$^{14}$, Bin Yang$^{15}$, Adrien Malgoyre$^{16}$, Colas Francois$^{17}$, Pierre Vernazza$^{18}$, Pierre Bourget$^{19}$, Emmanuel Jehin$^{20}$ and Alain Klotz$^{21}$.}

\date{\today}

\address{$^1\,$ Departamento de Ingeniería Eléctrica, Universidad de Chile,
$^2\,$ SYRTE, Observatoire de Paris, Université PSL, CNRS, Sorbonne Université,
$^3\,$ LFCA, CNRS,
$^4\,$ Departamento de Astronomía, Universidad de Chile,
$^5\,$ Facultad de Ingeniería y Ciencias, Universidad Adolfo Ibáñez,
$^6\,$ Millennium Institute for Astrophysics,
$^7\,$ Instituto de Astronomía y Ciencias Planetarias, Universidad de Atacama,
$^8\,$ Departamento de Ciencias Geológicas, Universidad Católica del Norte,
$^{9\,}$ Data Observatory Foundation,
$^{10\,}$ Centro de Astronomía (CITEVA), Universidad de Antofagasta,
$^{11\,}$ CASSACA,
$^{12\,}$ Instituto Multidisciplinario de Investigación y Postgrado - Departamento de Astronomía, Universidad de La Serena,
$^{13\,}$ Universidad de Valparaíso, Instituto de Física y Astronomía,
$^{14\,}$ University of Mendoza and CONICET,
$^{15\,}$ Instituto de Estudios Astrofísicos, Facultad de Ingeniería y Ciencias, Universidad Diego Portales, 
$^{16\,}$ Aix-Marseille University, CNRS, OSU-Pytheas,
$^{17\,}$ IMCCE, Observatoire de Paris, CNRS UMRO 8028, PSL Research University,
$^{18\,}$ Aix-Marseille Univ., CNRS, CNES, Laboratoire d’Astrophysique de Marseille,
$^{19\,}$ European Southern Observatory,
$^{20\,}$ Space Sciences, Technologies and Astrophysics Research Institute, Université de Liège,
$^{21\,}$ IRAP, Université de Toulouse, CNRS, UPS.}
\email{fripon.contact@das.uchile.cl }
\maketitle

\let\thefootnote\relax
\footnotetext{IMC2023}

\begin{abstract}
\href{https://www.fripon.org/}{{\color{blue} FRIPON}} is an efficient ground-based network for the detection and characterization of
fireballs, which was initiated in France in 2016 with over one hundred cameras, and which has been very successfully extended to Europe and Canada with one hundred more stations. After seven successful years of operation in the northern hemisphere, it seems necessary to extend this network towards the southern hemisphere - where the lack of detection is evident - to obtain an exhaustive view of fireball activity.
The task of extending the network to any region outside the northern hemisphere presents the challenge of
a new installation process, where the recommended and tested versions of the several sub-systems that
compose a station had to be replaced due to regional availability and compatibility considerations, as well
as due to constant software and hardware obsolescence and updates.
In Chile, we have a unique geography, with a vast extension in latitude, as well as desert regions, which
have generated the need to evaluate the scientific and technical performance of the network under special
conditions, prioritizing the optimization of a set of factors related to the deployment process, as well as
the feasible and achievable versions of the required components, the geographical location of the stations,
and their respective operational, maintenance, safety, and communication conditions.
In this talk, we present the current status of this effort, including a brief report on the obstacles and
difficulties encountered  and how we have solved them, the current operational status of the
network in Northern Chile, as well as the challenges and prospects for the densification of the network over
South America.
\end{abstract} 

\vspace{0.5cm}
{\LARGE{\textbf{Introduction}}}
\vspace{0.2cm}

The FRIPON project originates from the initiative of a team of French astronomers from various academic and research institutions, with two main objectives: to determine the origin of the fireball, and the recovery of fresh meteorites. This is achieved through a network of all-sky cameras that record the entire sky, 24 hours a day. The detection of fireballs by two or more cameras simultaneously enables the estimation of various parameters, primarily leading to the orbital parameters of the meteoroids, and a search radius for the fragment on the ground.

Up to this point, the network is concentrated in France and surrounding countries (for instance in Italy through the PRISMA network\cite{1}, and in Romania through the MOROI network\cite{2}). There are also about ten stations in Canada as well as some emerging installations in other parts of the world. Until now, almost all the cameras were located in the northern hemisphere, leaving the southern hemisphere without a history of detection and scientific output from the project.

Since Fireball registered events are not distributed uniformly across the Earth's surface \cite{3}, their global statistical analysis with the current FRIPON network will be biased. This is the main reason why expanding the network to the southern hemisphere is necessary.

In the southern hemisphere, Chile is a promising candidate due to several prerequisites for the expansion of the FRIPON network. It boasts good stability and internet connectivity throughout its territory, along with a reliable electrical grid. Geographically, northern Chile features an extensive desert with clear skies for most of the year, making it ideal for astronomy in general.

Considering the above, the following section describes the current state of the FRIPON network in Chile, followed by the hardware and software issues of the stations, and the installation challenges in the network expansion process. Finally, the future development of the FRIPON network in Chile is presented.

\vspace{0.5cm}
{\LARGE{\textbf{Current status of FRIPON-Chile}}}
\vspace{0.2cm}

During the year 2023, there has been a concentrated and progressive installation of new detection stations for the FRIPON network in Chile by a group of Chilean and French scientists (see the web page of \href{https://www.fcla.cl/fripon-chile}{{\color{blue} FRIPON-Chile}}). This new array of cameras constitutes the primary detection region in the southern hemisphere to date, as eight stations have been established covering the central and northern areas of the country, with an additional nine stations planned in the short term.

The hardware components that constitute a station, as well as the tasks of production, installation, and maintenance, require funding from institutions or individuals interested in contributing a station to the network – these are known as financiers. Additionally, the installation site where the station is housed requires a location that provides the necessary support, such as internet and electrical connectivity, security, and a person responsible for performing basic maintenance tasks if needed. This individual maintains communication with the technical team of the FRIPON network and is referred to as the host.

In Chile, the financiers include the Millennium Institute of Astrophysics (MAS) for the northernmost four cameras, FRIPON-France (at Paranal and La Silla), CASSACA (Salamanca station), ObsTech (El Sauce station), and the Franco-Chilean Laboratory of Astronomy (FCLA) providing an additional six cameras.

\begin{figure}[htb]
\centering
\includegraphics[width = 8cm]{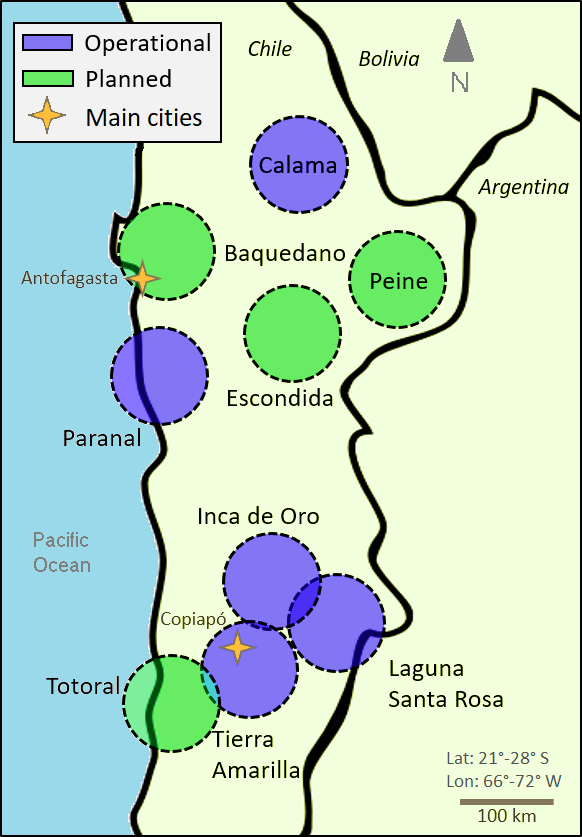}%
\vspace*{3pt}%
\caption{Map of the northern region of Chile displaying the distribution of operational and planned detection stations up until early September 2023.}
\label{2023-L12-sheep-figure1}
\end{figure}

Starting from the north, the Calama station is installed on the roof of the Chuquicamata School located south of the city of Calama at an altitude of around 2,260 meters (see Fig. \ref{2023-L12-sheep-figure1}). This is the northernmost operational FRIPON station in the country. The Paranal station is situated at the Paranal Observatory (ESO). The camera is installed in the dome of the TRAPPIST telescope\cite{4} and was the first station installed in Chile and the only one in the country until last year. There are three more planned stations between Paranal and Calama: Baquedano, Peine, and Escondida.

In the Atacama region, there are three operational stations strategically positioned to form a triangle. These are Inca de Oro, hosted at the ``El Pirquén'' restaurant in the homonymous commune. This camera could not be installed at the local observatory due to security issues. Next is the Laguna Santa Rosa station, located near the Maricunga salt flat, hosted at the refuge of the same name. This station is unique as it operates in extreme conditions, with temperatures dropping to around -20°C during the night, and it's situated at an altitude of 3750 meters above sea level, the highest FRIPON camera in the world.. The third operational station in this region is Tierra Amarilla, hosted in a Coya indigenous farm community, at an astrotourism site called ``Desierto Cósmico''. A planned camera is set for Totoral, southwest of Copiapó, to establish the connection between stations in the central-northern and northern zones.

\begin{figure}[htb]
\centering
\includegraphics[width = 8cm]{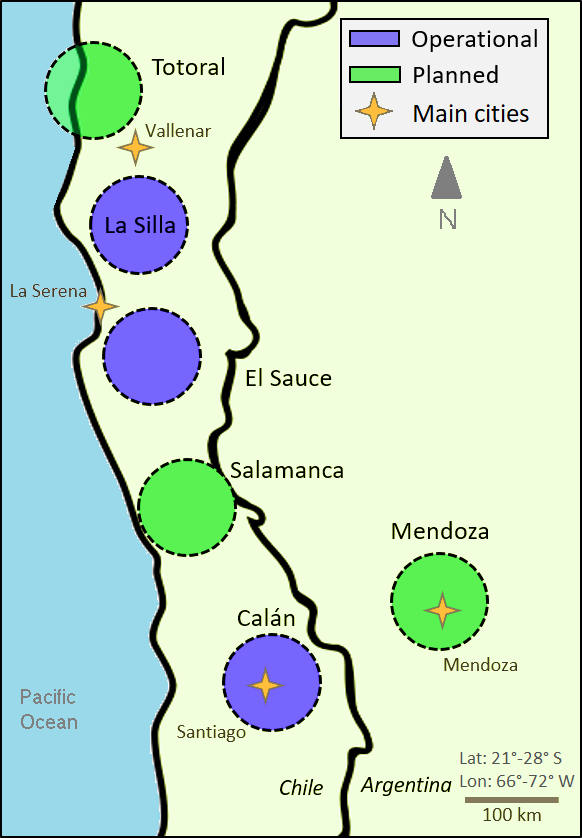}%
\vspace*{3pt}%
\caption{Map of the central-northern zone of Chile depicting the distribution of operational and planned detection stations as of early September 2023.}
\label{2023-L12-sheep-figure2}
\end{figure}

The first station in the central-northern zone is located at the La Silla Observatory (ESO), utilizing the facilities of the TAROT experiment\cite{5} (see Fig. \ref{2023-L12-sheep-figure2}). The next station is situated at an observatory dedicated to hosting telescopes, named Observatorio El Sauce (ObsTech), and it is the second camera installed in the country after Paranal. The southernmost operational station to date is located at the National Astronomical Observatory Cerro Cálan in the city of Santiago.

The installation of two more stations is in progress: one between El Sauce and Calán near Salamanca, and another to the east, in Argentina, located at the Technical School of Mendoza University south of the city of Mendoza. The stations are on average spaced about 150 km apart, with the closest ones around 90 km apart and the farthest separated by 250 km.

\newpage
\vspace{0.5cm}
{\LARGE{\textbf{Hardware and software issues}}}
\vspace{0.2cm}

The hardware components that make up a detection station are the camera, the computer, and the switch. Other elements are support components, like Ethernet cables, power cables, and adapters. Each of these devices requires its respective software for proper functioning.

Due to the constant updating and obsolescence of equipment and software, installations vary over time. This is due to the need to adapt the set of codes and standards designed for the FRIPON network stations to new equipment and operating systems, which may not always be compatible with the network's developments up to the creation date of a new station. Additionally, some equipment that once worked well with old configurations become obsolete over time and stops being manufactured. This demands finding ways to make new versions compatible and function optimally.

Starting with the computer, until recently, a NUC-type CPU, a compact and energy-efficient computer, was commonly used. However, the availability of these NUCs has been decreasing in the market. As a result, alternatives have been sought, and currently, computers from the ASUS Prime H510M-E brand are used. These new devices have a standard-sized case and higher power consumption than NUCs. Nevertheless, their advantage lies in their wide availability in the local market, facilitating their acquisition for future installations in the short and medium term.

To carry out an installation, the base operating system for the FRIPON system of a station is loaded onto a USB device. Then, the link to the specific and unique installation file for each station is introduced. However, an initial obstacle emerged: during the installation, the operating system did not recognize the built-in Ethernet card in the computer due to the lack of the appropriate driver. Despite attempts to manually install the driver, success was not achieved. An attempt was also made to use a USB-LAN adapter for internet connection, but although the installation was completed, the subsequent operation did not recognize this connection method. The solution to this problem involved installing new Ethernet cards in the computers of the TP-LINK T1500G-10PS brand and model, which connects to ports intended for additional components. The Debian version used in these new installations already includes the necessary drivers.

The second obstacle encountered when performing installations on these computers was the lack of compatibility with the version of the Debian operating system that was typically used to configure them. Through trial and error with newer versions of Debian, a version compatible with the computers and the FRIPON system was found, in this case, the Debian 10.3 operating system.

Each computer uses two storage disks: a 120GB SSD for the boot and processing files of the FRIPON system, and a 1000GB SATA disk for data storage. Problems arose in some installations due to the reversal of roles between the disks or because one of the disks had both functions while the other was empty. The station configuration in the LDAP platform (webpage to create, modify, and check stations parameters) allows assigning the correct function to each storage unit. However, in some cases, and for unknown reasons, the configuration was not successful, and it had to be corrected by running a directory change through SSH by the FRIPON technical team.

Regarding the switches, it was necessary to acquire them abroad as the required model was not available in Chile. Fortunately, the configuration file turned out to be compatible with these new models.


The cameras supported by the current FRIPON software are based on the model acA1300-30gm of Basler and the model 23G445 of DMK but the new FRIPON camera are now based on the new Basler model. To support these new kind of camera the FRIPON software has to be upgraded (probably by porting it to the 11 or 12 version of Debian) and this can not be done by the FRIPON-Chile group but only but the main FRIPON technical center.

\vspace{0.5cm}
{\LARGE{\textbf{Network installation difficulties}}}
\vspace{0.2cm}

Several factors present challenges for the successful expansion of the FRIPON network, including equipment and software updates and obsolescence, physical and environmental conditions at each location, as well as the availability of internet and electricity.

Stations in northern Chile face environmental conditions that can affect their performance, such as those typical of the Atacama Desert: intense solar radiation, wide day-night temperature variations and geographical isolation. For instance, the Calama station, located south of the city, is constantly exposed to dust deposition due to its proximity to the Chuquicamata copper mine. Additionally, the solar radiation is intense, requiring ongoing communication with the camera personnel at the local school to request dome cleaning tasks.

In the case of the cameras in the Atacama region, Ethernet cables are covered with PVC tubes to increase durability and protect them from the intense solar radiation. This is done even when the cables already have high-quality shielding and protection. Overall, all critical parts exposed to the exterior are installed with a higher level of protection against environmental harrasment.

A special case is the Laguna Santa Rosa station, situated near the Maricunga salt flat at an altitude of approximately 3750 meters above sea level. Here, nighttime temperatures can drop to -20°C, and the proximity to the salt flat creates a saline environment. The station is powered by solar energy and has satellite internet, although it faces frequent internet and electricity outages. To address this issue, the computer was replaced with a low-consumption one with an internal battery to mitigate the electrical load.

Another station with similar challenges is located on a farm in the mountains of Tierra Amarilla. This station is also powered by solar energy and expanded its battery capacity. Furthermore, the internet connection was switched from weak 4G in that area to Starlink satellite internet.

Chile's unique geography, a long and narrow strip in latitude, has resulted in a network of stations with a linear distribution, mainly in the central, central-northern, and central-southern zones. There are plans to install stations in the Argentine territory on the other side of the Andes mountain range.

Despite these limitations, stations are being planned in the southernmost region of Chile, in Tierra del Fuego (Patagonia), despite its cloudy and unstable climate. The geography of this area could facilitate the recovery of meteorites considering that much of the region has low vegetation.

\vspace{0.5cm}
{\LARGE{\textbf{Future of the Chilean FRIPON network development}}}
\vspace{0.2cm}

The following installations consider eight new stations throughout Chile and one in Argentina.

In the north, we will complete the network that will cover a large part of the Atacama Desert, three stations will be added between the ones already operating in Calama and Paranal. These are Baquedano, Escondida, and Peine. Each of them will be approximately 130 km apart from each other.

Further south, there's a group of three stations in the Atacama region, which will be connected to those located further south through the Totoral station.

Closer to the capital of Chile, the Salamanca station will be located to cover the long stretch of almost 400 km between the Calán and El Sauce stations.

Due to the narrowing of the territory towards the center of the country, a station will be installed near East of Santiago, in the city of Mendoza in Argentina, in order to broaden the linear distribution that extends throughout Chile.

To extend the reach of the network, the installation of three more cameras are planned before the end of 2024 in the Magallanes region, in the cities of Puerto Natales, Punta Arenas, and Puerto Williams, one of the southernmost cities in the world. These cameras will allow to survey areas at extreme southern latitudes, where no fireball detection network has reached before.

\vspace{0.5cm}
{\LARGE{\textbf{Conclusion}}}
\vspace{0.2cm}

Despite the difficulties described, we succeeded in installing eight FRIPON stations in Chile this year, between Calama and Santiago. These cameras are operational and last month we obtained the first events (five multiple detections with a subset of six cameras). Nine others cameras will be installed before the end of 2024, creating an efficient network in the Northern, central-northern and the patagonian region of Chile.

\vspace{0.5cm}
{\LARGE{\textbf{Acknowledgements}}}
\vspace{0.2cm}

FGR, SB and RAM acknowledge support form the French-Chilean Laboratory for Astrophysics (FCLA). RAM acknowledges partial funding from the Vicerrectoria de Investigacion y Desarrollo (VID) of the Universidad de Chile, project ENL02/23. MTP acknowledges the support of the Fondecyt-ANID Post-doctoral fellowship no. 3210253.

\end{document}